\documentclass[a4paper,dvipdfmx]{iopart}
\usepackage{graphicx}
\usepackage{color}
\usepackage{iopams}

\newcommand{\lrangle}[1]{\langle #1 \rangle}
\newcommand{\rmIA}{\mathrm{IA}}
\newcommand{\hrss}{h_\mathrm{rss}}

\begin{document}

\title[Amplitude-based detection method for GW bursts with HHT]%
{Amplitude-based detection method for gravitational wave bursts with the Hilbert-Huang Transform}

\author{Kazuki Sakai$^{1}$, Ken-ichi Oohara$^{2}$, Masato Kaneyama$^{3}$ and Hirotaka Takahashi$^{4}$}

\address{${}^{1}$Department of Information Science and Control Engineering,%
  Nagaoka University of Technology, Nagaoka, Niigata 940-2188, Japan}
\address{${}^{2}$Graduate School of Science and Technology,%
  Niigata University, Niigata 950-2181, Japan}
\address{${}^{3}$Department of Physics, Graduate School of Science,%
  Osaka City University, Osaka 558-8585, Japan}
\address{${}^{4}$Department of Information and Management Systems Engineering,%
  Nagaoka University of Technology, Nagaoka, Niigata 940-2188, Japan and \\%
  Earthquake Research Institute,%
  The University of Tokyo, Bunkyo-Ku, Tokyo 113-0032, Japan}

\ead{k\_sakai@stn.nagaokaut.ac.jp}

\begin{abstract}
  We propose a new detection method for gravitational wave bursts.
  It analyzes observed data with the Hilbert-Huang transform,
    which is an approach of time-frequency analysis
    constructed with the aim of manipulating non-linear and non-stationary data.
  Using the simulated time-series noise data
    and waveforms from rotating core-collapse supernovae at $30$ kpc,
    we performed simulation to evaluate the performance of our method
    and it revealed the total detection probability to be 0.94 without false alerms,
    which corresponds to the false alarm rate $< 10^{-3}$ Hz.
  The detection probability depends on the characteristics of the waveform,
    but it was found that
    the parameter determining the degree of differential rotation of the collapsing star
    is the most important for the performance of our method.
\end{abstract}


\vspace{2pc}

\submitto{\CQG}

\section{Introduction}
  The full operations of the advanced ground-based laser interferometer detectors for gravitational waves,
    such as Advanced LIGO~\cite{aLIGO2015} in the USA,
    Advanced Virgo~\cite{AdvVirgo2015} in Italy and KAGRA~\cite{KAGRA2013} in Japan,
    are about to be implemented.
  Their main targets are gravitational waves from compact binary coalescences,
    because waveforms of them are predictable with high accuracy
    in the post-Newtonian approximation of general relativity~\cite{ref:cbc}.
  In other words,
    gravitational waves from compact binary coalescences can be detected by means of matched filtering analysis.
  
  Gravitational wave bursts from various sources,
    such as gamma-ray bursts, core-collapse supernovae, soft-gamma repeaters,
    cosmic strings, late inspiral and mergers of compact binaries,
    ring-downs of perturbed neutron stars or black holes,
    are also thought as detectable by these detectors as well.
  However, it is not easy to predict waveforms from these sources
    with sufficient high accuracy for matched filtering,
    because their dynamics are complicated and the equation of state of neutron star matter is not known so well.
  Thus, a detection method that does not require exact information of waveforms
    is needed for the detection of gravitational wave bursts.
  Anderson {\it et al.}~\cite{AndersonPRD2001} have given a solution to it,
    namely, the Excess Power Method.
  It analyzes an observed data in time-frequency space by means of the short time Fourier transform.
  It requires only knowledge of the duration and frequency band of target signals.
  Thus, the Excess Power Method is well accepted in the field of gravitational wave data analysis.
  Recently, several new approaches for the detection of gravitational wave bursts
    with no or a little assumptions on the source models
    have been developed~\cite{ref:klimenko2008, ref:sutton2010, ref:thrane2011, ref:lynch2015, ref:klimenko2015, ref:cornish2015}.
  
  In fact, it is by burst search pipeline
    that the gravitational wave signal GW150914 was first identified~\cite{ref:LIGOPRL2016,ref:LIGOburst2016}.
  This event was announced by distinct search pipelines, coherent waveburst algorithm~\cite{ref:klimenko2008},
    omicron-LALInference-Bursts~\cite{ref:lynch2015}, and BayesWave~\cite{ref:cornish2015}.
  As shown there, in order to enhance credibility of a detected event,
    it is important to set up multiple pipelines designed to detect gravitational wave bursts.
  
  In this paper, we propose a new approach for amplitude-based detection method
    of gravitational wave bursts by using Hilbert-Huang transform (HHT).
  In 1998, Huang {\it et al.}\cite{HuangPRSA1998} proposed the HHT as a new approach of time-frequency analysis.
  Its characteristic feature is that it adaptively determines basis for decomposing according to the input data.
  The adaptiveness provides two specific merits.
  One is that it is capable of manipulating non-linear and non-stationary data,
    and the other is that it is not restricted by the uncertain principle.
  Since gravitational wave signals are mostly non-stationary and non-linear,
    the HHT is likely to be a powerful tool for analyzing gravitational wave signals.
  So far, it has been applied to some data analysis for gravitational waves~\cite{Jordan2007, StroeerPRD2009, Takahashi2013, ref:kaneyama_sub}.
  
  The HHT consists of two steps.
  One is the empirical mode decomposition (EMD) and the other is Hilbert spectral analysis (HSA).
  The EMD decomposes time-series data into finite numbers of intrinsic mode functions (IMFs),
    each of which represents simple intrinsic oscillation contained in the original data.
  The Hilbert transform is then performed on each IMF to obtain instantaneous amplitude (IA)
    and instantaneous frequency (IF).
  Consequently, we can extract time-evolutionary information of the data with respect to the amplitude and frequency.
  The EMD works as wavelet-like filter bank~\cite{FlandrinIEEE2004},
    and then noise will be dispersed into all IMFs with nearly equal amplitude.
  A burst signal of gravitational waves, on the other hand,
    will be decomposed into a single or a few set of IMFs,
    if it has a single oscillation mode.
  The IA of the burst signal is thus significantly large only at specific regions in some IMFs.
  Our method searches for the excess of the amplitude at these regions to know
    whether the data contains gravitational waves.
  
  To evaluate the performance of our method,
    we conduct a Monte Carlo simulation.
  An ensemble of input data is made
    by adding a colored Gaussian noise based on Advanced LIGO sensitivity curve~\cite{ref:aligonoisecurve}
    to a model signal of gravitational waves from a rotating core-collapse supernova
    as well as a sine-Gaussian signal.
  
  Throughout this paper, discrete sequences are represented with brackets, such as $x[n]$,
    and continuous functions are represented with parentheses, such as $s(t)$.

  This paper is organized as follows.
  In \sref{sec:data_set},
    we briefly describe simulated time-series data of detector and the waveforms that we employ in this paper.
  In \sref{sec:method},
    we explain the our proposed amplitude-based detection method with HHT.
  In \sref{sec:simulation} and \sref{sec:discussion},
    the results of the analysis with our proposed method are shown
    and the implication of the results is discussed.
  \Sref{sec:summary} is devoted to a summary.

\section{Data set}\label{sec:data_set}
\begin{figure}[tbp]
  \begin{center}
    \input{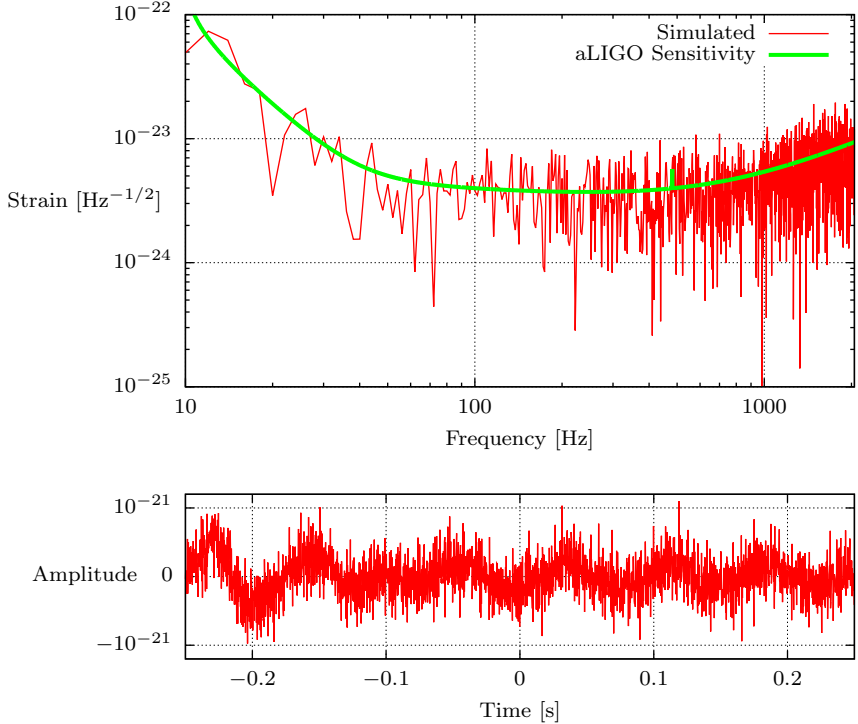}
  \end{center}
  \caption{Sensitivity curve of Advanced LIGO (green) and an example of simulated noise strain (red) in the upper panel,
        and corresponding time-series $n(t)$ in the lower panel.}
  \label{fig:NoiseSpectrum}
\end{figure}
  We use a sensitivity curve of Advanced LIGO, namely,
    the zero-detuned, high-power sensitivity curve~\cite{ref:aligonoisecurve},
    and produce 1000 realizations of Gaussian noise in frequency domain.
  Each of time-series noise data, $n(t)$,
    is produced by applying the inverse Fourier transform to the corresponding realization.
  The sampling frequency is 4096 Hz and the duration of each segment of the time series is 0.5 s.
  We set cut-off frequency at 10 Hz.
  \Fref{fig:NoiseSpectrum} shows the sensitivity curve and an example of simulated noise strain in the upper panel,
  as well as the corresponding time-series $n(t)$ in the lower panel.

  Concerning the injected signals,
    we first consider sine-Gaussian signals as simple models of short-duration gravitational-wave transients.
  As shown later,
    the gravitational wave signal from a core-collapse supernova
      is mostly characterized by a peak at the time of the core bounce
    and it can be emulated by a sine-Gaussian signal.
  Since the sine-Gaussian signal is narrowband,
    we can test the performance of detection methods in those frequency bands.
  The sine-Gaussian signal that we used is expressed as
  \begin{eqnarray}
    h(t) = \alpha \exp \left( - \frac{2 \pi^2 f^2 t^2}{Q^2} \right) \cos (2 \pi f t),
    \label{eq:SineGaussian}
  \end{eqnarray}
  where $f$, $Q$ and $\alpha$ are constants.
  The amplitude $\alpha$ will be determined
    such that $\hrss$ of each signal was equivalent to the average of $\hrss$ of simulated burst signals,
    where $\hrss$ is defined by
    \begin{eqnarray}
      \hrss = \sqrt{ \int_{-\infty}^{\infty} \rmd t \, |h(t)|^2 }\, .
    \end{eqnarray}
  
  Next, as more realistic models,
    we use waveforms from rotating core-collapse supernovae given by Dimmelmeier {\it et al.}~\cite{Dimmelmeier2008}.
            
  In the simulation by Dimmelmeier {\it et al.},
    they used two nonzero-temperature equations of state (EoS),
    to simulate some thermofluid dynamic quantities.
  One is by Shen {\it et al.}~\cite{ShenPTP1998} (Shen EoS),
    the other is by Lattimer and Swesty~\cite{LattimerNPA1991} (LS EoS).
  LS EoS is based on a compressible liquid-drop model,
    and Shen EoS is based on a relativistic mean field model.
  As presupernova stellar models,
    they employed various nonrotating models with zero-age main sequence mass
      $M_{\rm prog} = 11.2 M_{\odot}$, $15.0 M_{\odot}$, $20 M_{\odot}$ and $40 M_{\odot}$
      (core-models s11, s15, s20 and s40, respectively),
    and some rotating models with
      $M_{\rm prog} = 15.0 M_{\odot}$ (core-models e15a and e15b)
      and $20.0 M_{\odot}$ (core-models e20a and e20b).
  They set those cores that are initially nonrotating (core-models s11, s15, s20, and s40) artificially into rotation,
    whose specific angular momentum $j$ is given by
    \begin{eqnarray}
        j = A^2 (\mathit{\Omega}_{\rm c,i} - \mathit{\Omega}).
    \end{eqnarray}
  Here $A$ is the length that parameterizes the degree of differential rotation,
    $\mathit{\Omega}$ is the angular velocity,
    and $\mathit{\Omega}_{\rm c,i}$ is the precollapse value of $\mathit{\Omega}$ at the center.
  The differentiality becomes stronger as $A$ is decreasing.
  In the Newtonian limit, this reduces to
    \begin{eqnarray}
        \mathit{\Omega} = \mathit{\Omega}_{\rm c,i} \frac{A^2}{A^2 + r^2 \sin^2 \theta} ,
    \end{eqnarray}
    where $r \sin \theta$ is the distance to the rotating axis.
  The precollapse rotation is parameterized by in terms of $A$ and $\mathit{\Omega}_{\rm c,i}$,
    where $A = 50\,000$ km indicates an almost uniform rotation,
    $A = 1000$ km indicates a moderately differential rotation,
    and $A = 500$ km indicates a strongly differential rotation.

\begin{figure}[tbp]
    \centering
    \input{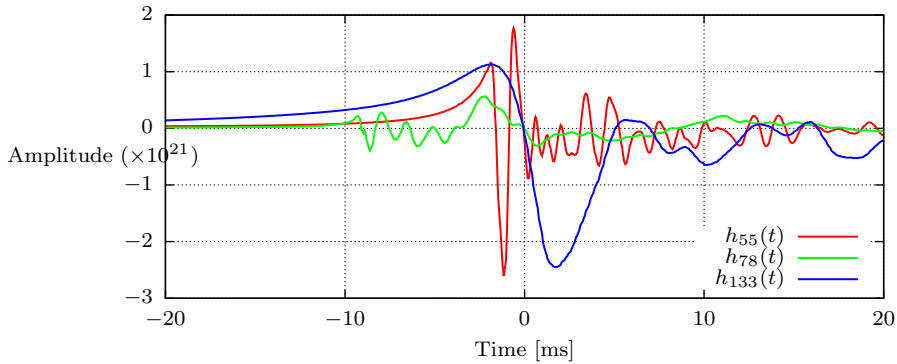}
    \caption{Examples of the simulated waveforms. 
             The waveforms are numbered in alphabetical order of their waveform names.
             The parameters of them are listed in \tref{tab:exParameters}.}
    \label{fig:waveforms}
  \end{figure}

  In \cite{Dimmelmeier2008},
    the core collapse models are labeled
     according to $M_{\rm prog}$, $A$ and $\mathit{\Omega}_{\rm c,i}$.
  We sort them alphabetically by the original labels and assign waveform numbers to them,
    that is, the waveforms in this paper are sorted by the values of
      $M_{\rm prog}$, $A$, $\mathit{\Omega}_{\rm c,i}$, and then the EoS.
  As examples, the waveforms of 55, 78 and 133 are plotted in \fref{fig:waveforms}
    and their parameters are listed in \tref{tab:exParameters}.
  
  \begin{table}[tbp]
    \centering
    \caption{Parameters of the waveforms plotted in \fref{fig:waveforms}.
             The parameter in each column is the same as \tref{tab:Parameters} to be shown later.}
    \label{tab:exParameters}
    \lineup
    \begin{indented}
    \item[]
      \begin{tabular}{ccrcr}
      \br
        Waveform  &  Core Model  &  $A$ [km]  &  $\mathit{\Omega}_{\rm c,i}$ [rad/s]  &  EoS  \\
      \mr
          {\0}55  &       s15    &  1000      &                                 4.56  &  LS   \\
          {\0}78  &       s20    &  50\,000   &                                 1.43  &  Shen \\
             133  &       s40    &  500       &                                 2.71  &  LS   \\
      \br
      \end{tabular}
    \end{indented}
  \end{table}

\section{Detection Method}\label{sec:method}
  Our proposed method consists of a preprocessing part and an event trigger part.
  In the preprocessing part,
    the HHT is performed to whitened observed data.
  In the event trigger part,
    an event is triggered
    if an index value defined with respect to the amplitude of the preprocessed data
      exceeds a predetermined threshold.
  We call this method the {\it Excess Amplitude Method} (henceforth {\it EAM}).
  
  The whitening process of observed data is conducted
    by using a linear predictive error filter (LPEF).
  The LPEF is one of finite impulse response filters
    that flatten the spectrum of data~\cite{TheoryOfLinearPrediction}.
  Cuoco {\it et al.}~\cite{CuocoCQG2001} have examined
    the availability of the LPEF as a whitening method for signals from gravitational wave detectors,
    and it also has been incorporated in some gravitational wave search methods~\cite{McNabbCQG2004,ChatterjiCQG2004}.
  Refer to the review by Chatterji {\it et al.}~\cite{ChatterjiCQG2004} for details.
  
\begin{figure}[tbp]
    \centering
    \input{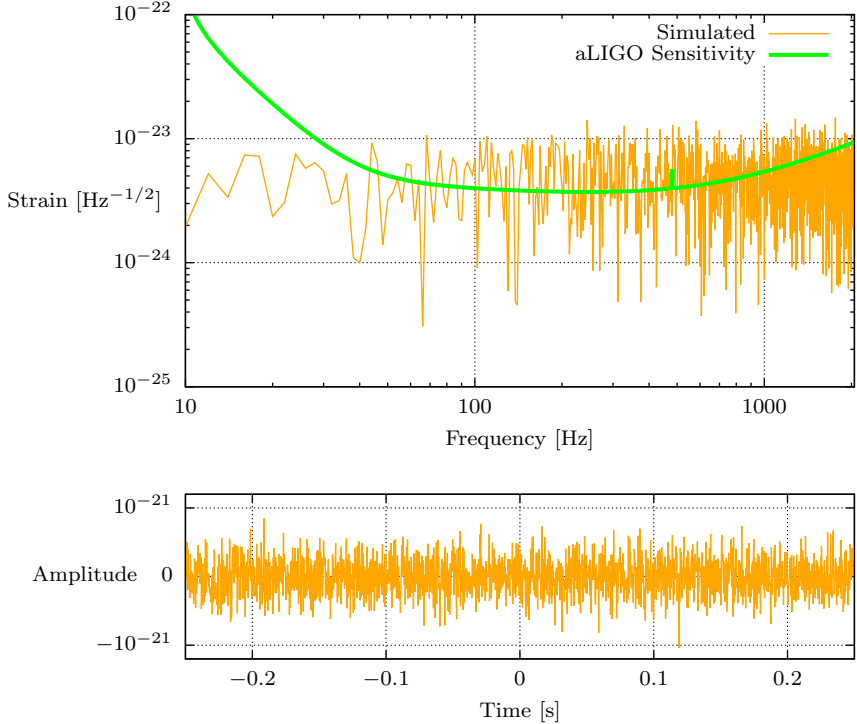}
    \caption{The result of whitening of the simulated noise data shown in \fref{fig:NoiseSpectrum}
               by means of the LPEF.
             The spectrum of the whitened data as well as the sensitivity curve of Advanced LIGO
               is plotted in the upper panel
               and the time-series data,
                 that is, the inverse Fourier transform is shown in the lower panel.}
    \label{fig:whiteningTest}
  \end{figure}

  \Fref{fig:whiteningTest} shows the result of the LPEF
    applying to the simulated noise data of Advanced LIGO (\fref{fig:NoiseSpectrum}).
  We tuned the filter by using 1000 patterns of simulated noise data.
  
  The HHT is a method of time-frequency analysis for time-series data,
    developed by Huang {\it et al.}~\cite{HuangPRSA1998}.
  The HHT consists of two steps,
    the empirical mode decomposition (EMD) and the Hilbert spectral analysis (HSA).
  The HSA is based on the analytic representation of a signal $s(t)$~\cite{Cohen1995} such as
    \begin{eqnarray}
        z(t) &= s(t) + \rmi \mathcal{H}[s(t)]  \nonumber \\
             &= a(t) \exp \left( \rmi \theta(t) \right),
    \end{eqnarray}
    where $\mathcal{H}[s(t)]$ is the Hilbert transform of $s(t)$ defined by
    \begin{eqnarray}
        \mathcal{H}[s(t)] = \frac{1}{\pi} \mathrm{PV} \int_{-\infty}^{\infty} \rmd t' \, \frac{s(t')}{t - t'},
        \label{eq:HilbertTransform}
    \end{eqnarray}
    where PV denotes the Cauchy principle value.
  Then, the instantaneous amplitude (IA) $a(t)$ and the instantaneous phase $\theta(t)$ are given by
    \begin{eqnarray}
        a(t) = \sqrt{s(t)^2 + (\mathcal{H}[s(t)])^2}
        \quad \mbox{and} \quad
        \theta(t) = \arctan \left( \frac{\mathcal{H}[s(t)]}{s(t)} \right),
        \label{eq:IAandIPhase}
    \end{eqnarray}
    respectively.
  The instantaneous frequency (IF) $f(t)$ is defined as the derivative of $\theta(t)$;
    \begin{eqnarray}
        f(t) = \frac{1}{2 \pi} \frac{\rmd \theta(t)}{\rmd t}.
        \label{eq:IF}
    \end{eqnarray}

  Unfortunately, however,
    the HSA will not lead physically meaningful results
      unless the signal $s(t)$ is ``monocomponent'' and ``narrowband''~\cite{HuangPRSA1998,Cohen1995}.
  The basic role of the EMD is to decompose time-series data into finite numbers of IMFs,
    each of which represents a simple intrinsic oscillation mode.
  The IMF has two specific properties, namely,
    the numbers of its extrema and zero-crossing are the same or differ by one,
    and its local mean is zero at any point.
  Owing to these properties,
    the IMF admits a well-behaved Hilbert transform.
  The original signal $s(t)$ is sum of IMFs $c_i(t)$ and residual $r(t)$ of the EMD,
    \begin{eqnarray}
        s(t) = \sum_{i=1}^{N_{\rm IMF}} c_i(t) + r(t).
    \end{eqnarray}
    where $N_{\rm IMF}$ is the number of IMFs.
  
  Performing the Hilbert transform on the IMFs,
    the HHT tells multiple IAs and IFs
    if the signal $s(t)$ contains multiple oscillations and/or a broadband noise.
  
  Because the HHT is not based on the assumption that the signal is stationary or linear,
    it has a potential to analyze non-stationary and non-linear data.
  Detailed descriptions of HHT algorithm used in this paper are found in~\cite{Takahashi2013,ref:kaneyama_sub}.
  
  Since EMD works as wavelet-like filter bank~\cite{FlandrinIEEE2004},
    whitened noise whose spectrum distributes uniformly will be dispersed into almost all IMFs
      with nearly equal amplitude.
  A gravitational wave burst signal, on the other hand,
    will be mostly decomposed into a small number of IMFs,
    and hence the IAs of these IMFs become significantly large in some region.
  We can therefore expect that gravitational-wave bursts will be detected
    by finding the region where IA exceeds a certain threshold.
  
  As the first step of the event trigger part,
    we need to study the statistical characteristics of IA for the background,
    that is, the average $\mu_i$ and the standard deviation $\sigma_i$ of $\rmIA_i$
      for IMF$_i$ $(i=1,2,\dots,N_\mathrm{IMF})$.
  In the present simulation, $\mu_i$ and $\sigma_i$ are estimated
    from over 1000 realizations of whitened noise-only data.
  We then define the conditional averaging function $A_{k}(a_i[n] | \mu_i, \sigma_i)$
    and the maximum mean amplitude $x_k$, respectively, by
    \begin{eqnarray}
        \label{eq:CAFk}
        A_{k}(a_i[n] | \mu_i, \sigma_i) = %
        \cases{
          \lrangle{a_i[n]}_{k}  &  $(\lrangle{a_i[n]}_{k} \geq \mu_i + 4 \sigma_i)$,  \\
          0            &  otherwise,  \\
        }
        \\
        \lrangle{a_i[n]}_{k} = \frac{1}{k} \sum_{l=n}^{n+k-1} a_i[l],
    \end{eqnarray}
    where $a_i[n]$ is the value of $\rmIA_i$ at $t = t_n$ and $k$ is a predetermined window size,
    and
    \begin{eqnarray}
        x_k = \max_{1 \leq i \leq N_\mathrm{IMF}} \left(
                  \max_{1 \leq n \leq L_\mathrm{d} - k} A_{k}(a_i[n] | \mu_i, \sigma_i)
                \right), 
    \end{eqnarray}
    where $L_\mathrm{d}$ is the data length.
  The definition \eref{eq:CAFk} means
    that the input data is judged to contain no signal
    in the region if the IA is less than $\mu + 4 \sigma$ of the noise level.
  
  The EAM will trigger an event of a gravitational wave,
    if the index value $x_k$ is greater than or equal to a predetermined threshold value $x_\mathrm{TH}$.
  We will discuss appropriate values
    of the window size $k$ and the threshold $x_\mathrm{TH}$ in the next section.

\section{Simulation}\label{sec:simulation}
  To estimate the optimal window size $k$ and the optimal threshold $x_\mathrm{TH}$,
    we have conducted simulations of applying the EAM to simulated data.
  The data were superpositions of simulated gravitational wave burst signals
    (sine-Gaussian and core-collapse supernova signal)
    and each of 1000 realizations of simulated Advanced LIGO noise data.
  In this paper,
    we assumed that gravitational wave bursts had entered with optimal orientation of the detector
    and the event of emitting gravitational wave had occurred in our galaxy
    (at most 30 kpc apart from the earth).
  In this case,
    average $h_\mathrm{rss}$ of the simulated signals is $9.578 \times 10^{-21}$.
  
  To estimate false alerm rate (FAR),
    we also applied the EAM to each realization of simulated noise data without signal injection.
  
\begin{figure}[tbp]
    \centering
    \input{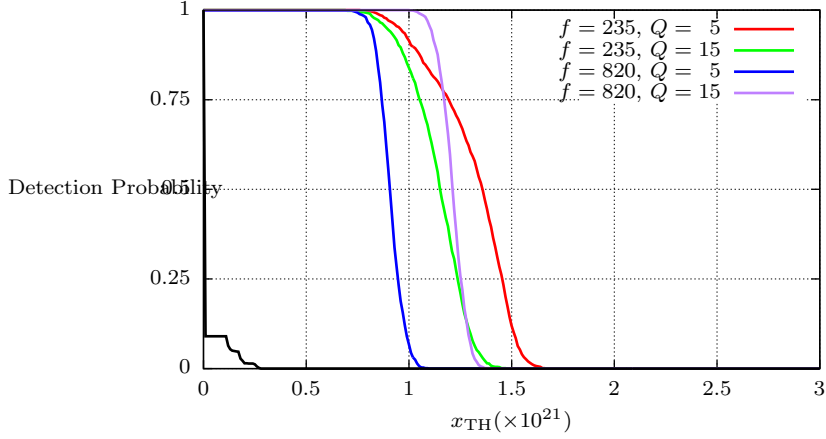}
    \caption{Detection probability against threshold value $x_\mathrm{TH}$ of the EAM
               to the sine-Gaussian signals defined by \eref{eq:SineGaussian}.
             The window size $k$ is set at $41$,
               which corresponds to the duration of about $10$ ms.
             The solid black line shows the false alerm probability of the EAM.}
    \label{fig:SGEAM41}
  \end{figure}

  First,
    we used the sine-Gaussian signals defined in \eref{eq:SineGaussian}.
  We considered the four cases:
    ($f=235$ Hz, $Q=5$), ($f=235$ Hz, $Q=15$), ($f=820$ Hz, $Q=5$) and ($f=820$ Hz, $Q=15$).
  We provisionally let the window size $k$ be $41$;
    this value means the duration of window is about 10 ms.
  \Fref{fig:SGEAM41} shows the variations in detection probability of sine-Gaussian signals
    in accordance with threshold value $x_\mathrm{TH}$.
  It reveals that the EAM will detect almost all events,
    namely the detection probability being at least 0.999,
    without false alerms
    when the threshold value is between $2.8 \times 10^{-22}$ to $7.2 \times 10^{-22}$.
  
\begin{figure}[tbp]
    \centering
    \input{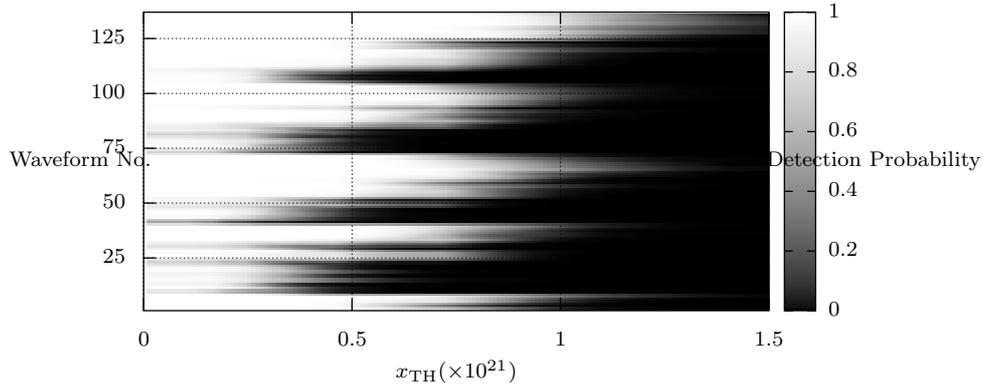}
    \caption{Detection probability against threshold value $x_\mathrm{TH}$ of the EAM
               to the simulated waveforms of core-collapse supernovae~\cite{Dimmelmeier2008}.
             The window size $k$ is the same as \Fref{fig:SGEAM41}.
             The waveforms are numbered in alphabetical order.}
    \label{fig:SNEAM41}
  \end{figure}

  Second,
    we used the simulated waveforms of core-collapse supernovae.
  \Fref{fig:SNEAM41} shows the variations in detection probability of them with threshold value.
  Detection probabilities vary with waveform,
    which corresponds to the EoS and the parameters of initial configurations of core-collapse models.
  For more than half waveforms,
    some values of threshold will give the detection probability of 1.0 without false alerms,
    while the detection probability never reaches 1.0 with any threshold value for some waveforms.
  
\begin{figure}[tbp]
    \centering
    \input{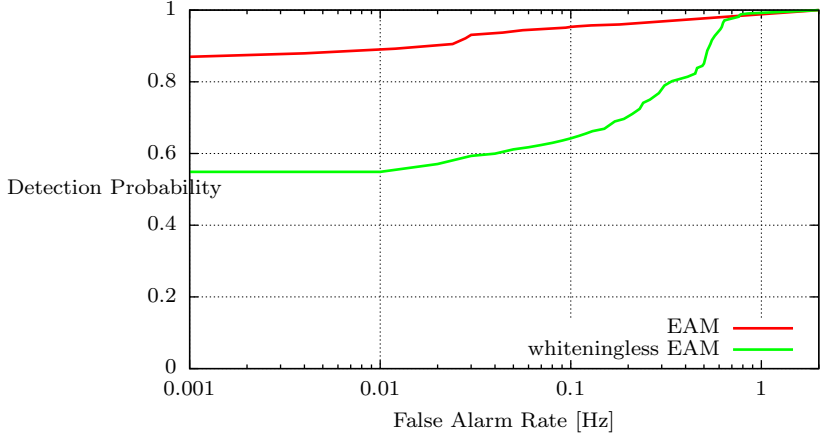}
    \caption{ROC curves of the EAM for all waveforms of core-collapse.
             The red and green lines show the ROC curves of the EAM with and without whitening, respectively.
             The window size $k$ is set at $41$ for the red line,
               while it is at $10$ for the green line,
               since these are respectively the optimal values.}
    \label{fig:SNROC41}
  \end{figure}

  To evaluate the performance of the EAM more comprehensively,
    we plotted a receiver operating characteristic (ROC) curve,
    which is created by plotting the detection probability of a target signal against the FAR.
  \Fref{fig:SNROC41} shows the ROC curve of the EAM for all waveforms of core-collapse we used.
  It reveals that the EAM has the detection probability of at least 0.85,
    with the false alarm rate $< 10^{-3}$ Hz.
  It also indicates the whitening process plays an important role to improve the efficiency in the EAM.
      
\begin{figure}[tbp]
    \centering
    \input{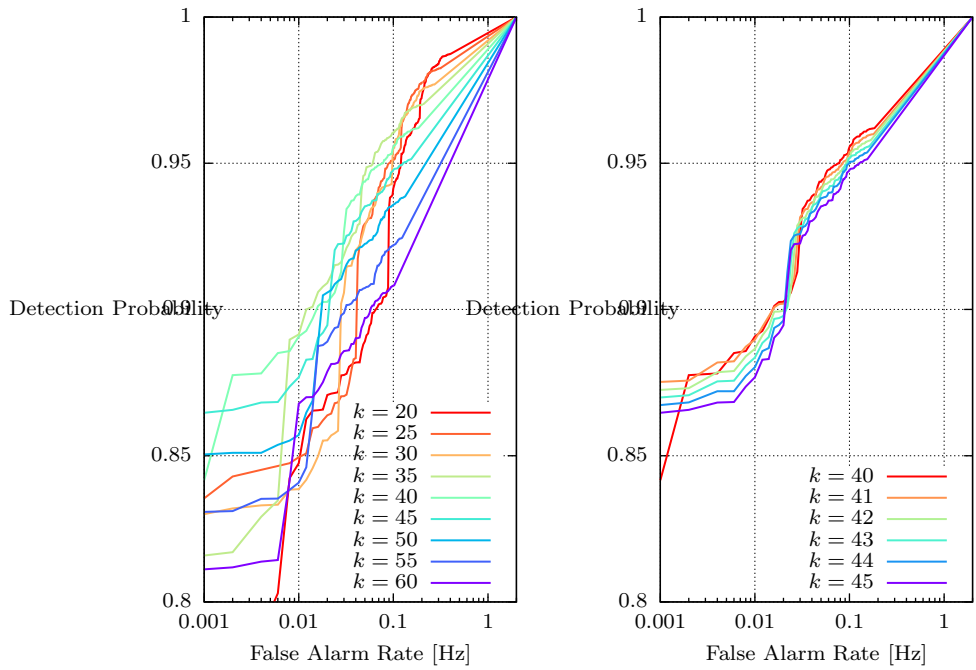}
    \caption{ROC curves of the EAM for all waveforms of core-collapse
               for various values of window size $k$.
             The left panel shows a comparison of the window size $k$ ranging from 20 to 60. 
             The right panel shows a minute comparison of $k$ ranging from 40 to 45.}
    \label{fig:SNROCtotal}
  \end{figure}

  We plotted the ROC curves of the EAM for various values of $k$ in \fref{fig:SNROCtotal}.
  A small window size is expected to give a high detection probability,
    since it can detect even short-term excess of the amplitude.
  However it is likely to lead a high FAR at the same time.
  Therefore,
    we have to look for the optimal value of the window size to the maximum of the detection probability
      with the minimum of the FAR.
  The left panel of \fref{fig:SNROCtotal} shows a comparison of the window size $k$ ranging from 20 to 60,
    and it indicates that the optimal value of $k$ lies between $40$ and $45$.
  A minute comparison in this range is made in the right panel of \fref{fig:SNROCtotal}
    and it suggests that the optimal value of $k$ is $41$.
  
\begin{figure}[tbp]
    \centering
    \input{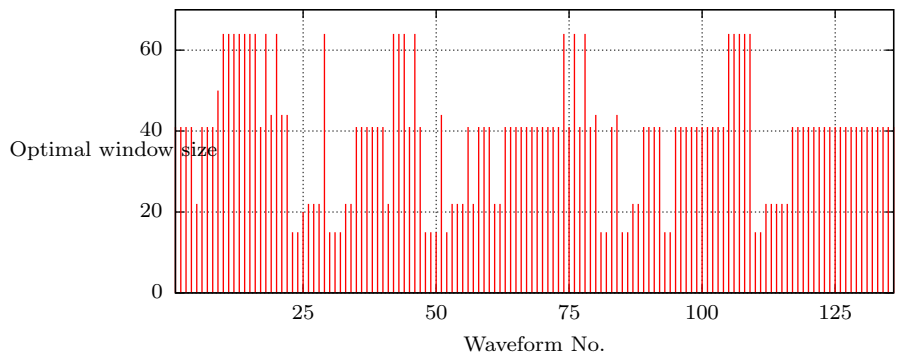}
    \caption{Optimal window size of each waveform.}
    \label{fig:SNkoptimal}
  \end{figure}

  We also checked the optimal window size for each waveform.
  If the detection probability of some waveforms becomes the highest value with two or more window sizes,
    the optimal window size of the waveform is chosen the one of them
      with which the total detection probability becomes highest.
  The optimal window size of each waveform is shown in \fref{fig:SNkoptimal}.
  The optimal value is 15, 22, 41 or 64 for all the waveforms with few exceptions.
  The exceptions are
    $k = 44$ for five waveforms, $k = 20$ for one waveform, and $k = 50$ for one waveform.
  
\begin{figure}[tbp]
    \centering
    \input{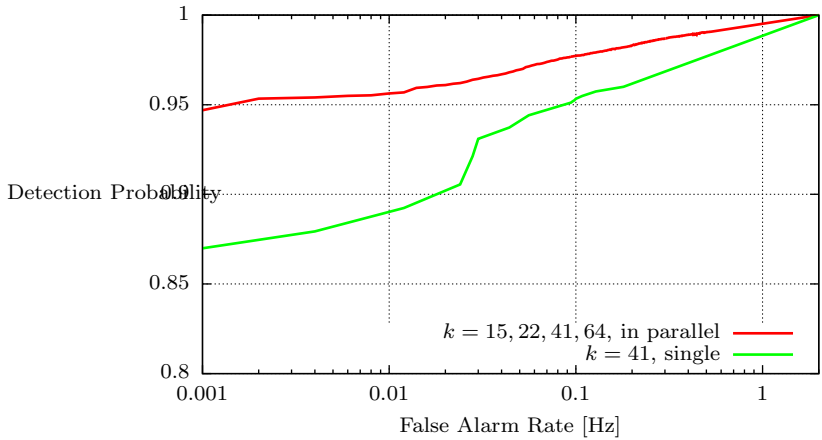}
    \caption{ROC curves of the EAM for all waveforms of core-collapse
             The red line is the result with window sizes $k=15,22,41,64$,
               while the green line is the result with window size $k=41$, only.}
    \label{fig:SNROCparallel}
  \end{figure}

  Consequently,
    we will perform the EAM analysis with these four values of $k$ simultaneously.
  Triggers are recorded if $x_k$ exceeds the threshold for at least one of $k$.
  The ROC curve of the EAM with some window size is shown in \fref{fig:SNROCparallel}.
  It reveals that the detection efficiency is improved significantly.
  Eventually,
    the detection efficiency
    (the detection probability without false alerms in our realizations,
      which corresponds to the false alarm rate $< 10^{-3}$ Hz in our simulation)
    of our proposed method reached 0.940.
  
\begin{figure}[tbp]
    \centering
    \input{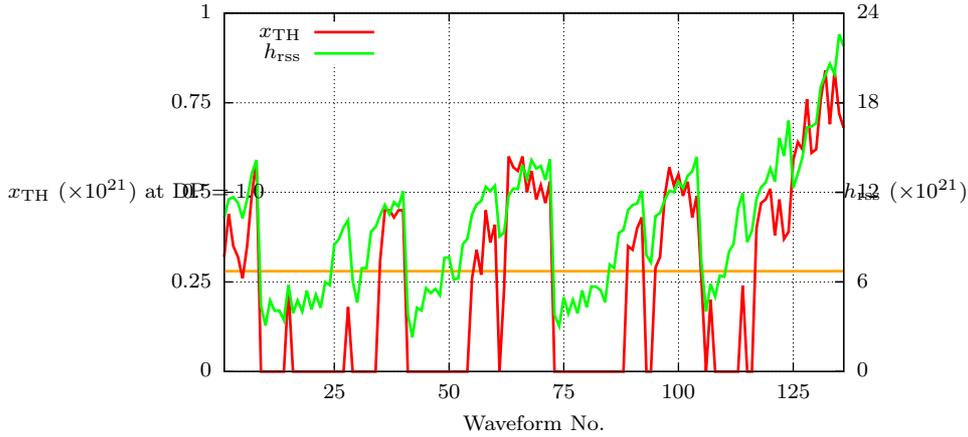}
    \caption{The threshold $x_\mathrm{TH}$ that makes the detection probability 1.0
               and the $h_\mathrm{rss}$ against the waveform number.
             The correlation coefficient is  $0.893$.
             The orange line represents $x_\mathrm{TH} = 0.28 \times 10^{-21}$, with which the FAR becomes 0.}
    \label{fig:SNhrssDP}
  \end{figure}

  As mentioned before with \fref{fig:SNEAM41},
    the threshold value that leads high detection probability depends on waveforms of its target signal.
  The detection probability can be 1.0 without false alerms for some waveforms
    (hereinafter referred to as class C1)
    if an appropriate threshold value $x_\mathrm{TH}$ depending on the waveforms is adopted,
    while it cannot be realized with any threshold values for other waveforms (class C2).
  In \fref{fig:SNhrssDP},
    we plotted $h_\mathrm{rss}$ for each waveform as well as the threshold values
      above which the detection probability becomes 1.0.
  It shows that there is a strong correlation between them,
    yielding the correlation coefficient of 0.893.
  These waveforms of class C1 have $h_\mathrm{rss}$ of $1.0 \times 10^{-20}$ or above,
    while only 4 of 66 waveforms of class C2 have $h_\mathrm{rss}$ of this value.
  The parameters of core-collapse simulations of class C1 are listed in \tref{tab:Parameters}.
  It indicates that the parameter $A$ of class C1 is 1000 km or below,
    while waveforms of $A=50\,000$km, which corresponds to almost uniform rotation, belong to class C2.
  As a result,
    the degree of differential rotation is more important for the EAM than the mass of the star.
  
  \begin{table}[tbp]
    \caption{Properties of the core-collapse models
               for which gravitational waves can be detected without false alerms by means of the EAM.
             The name of the core model derives from the rotation and mass of the presupernova model
               (see text in \sref{sec:data_set}).
             The parameters $A$ and $\mathit{\Omega}_{\mathrm{c,i}}$ denote, respectively,
               the differential rotation length scale and the precollapse angular velocity at the center of the star.
             As for the EoS,
               LS denotes the EoS of Lattimer and Swesty~\cite{LattimerNPA1991}
               and Shen denotes the EoS of Shen {\it et al.}~\cite{ShenPTP1998}.
             No gravitational wave signal corresponding to $A =50\,000$ km can be detected without false alerms.}
    \label{tab:Parameters}
    \lineup
    \begin{indented}
    \item[]
      \begin{tabular}{rccr@{}lc}
      \br
        Waveform    &  core-model  &  $A$ [km]  &  $\min \mathit{\Omega}_{\mathrm{c,i}}$  &  [rad/s]  &  EoS
      \\
      \mr
        1--2        &        e15   &      --    &                               {\0}4.18  &           &  LS/Shen
      \\
        7           &        e20   &      --    &                                  11.01  &           &  LS\phantom{/Shen}
      \\
        6, 8        &        e20   &      --    &                               {\0}3.13  &           &  \phantom{LS/}Shen\rlap{$^\dagger$}
      \\
        35--39      &        s11   &   {\0}500  &                                  10.65  &           &  LS/Shen
      \\
        59          &        s15   &      1000  &                               {\0}7.60  &           &  LS\phantom{/Shen}
      \\
        56, 58, 60  &        s15   &      1000  &                               {\0}4.56  &           &  \phantom{LS/}Shen\rlap{$^\dagger$}
      \\
        63--72      &        s15   &   {\0}500  &                               {\0}5.95  &           &  LS/Shen
      \\
        89--92      &        s20   &      1000  &                               {\0}6.45  &           &  LS/Shen
      \\
        95--104     &        s20   &   {\0}500  &                               {\0}5.95  &           &  LS/Shen
      \\
        117--124    &        s40   &      1000  &                               {\0}3.40  &           &  LS/Shen
      \\
        125--136    &        s40   &   {\0}500  &                               {\0}4.21  &           &  LS/Shen
      \\
      \br
      \end{tabular}
    \item[] $^\dagger$ The detection probability in LS EoS with the same parameters is 0.999.
    \end{indented}
  \end{table}
  
\begin{figure}[tbp]
    \centering
    \input{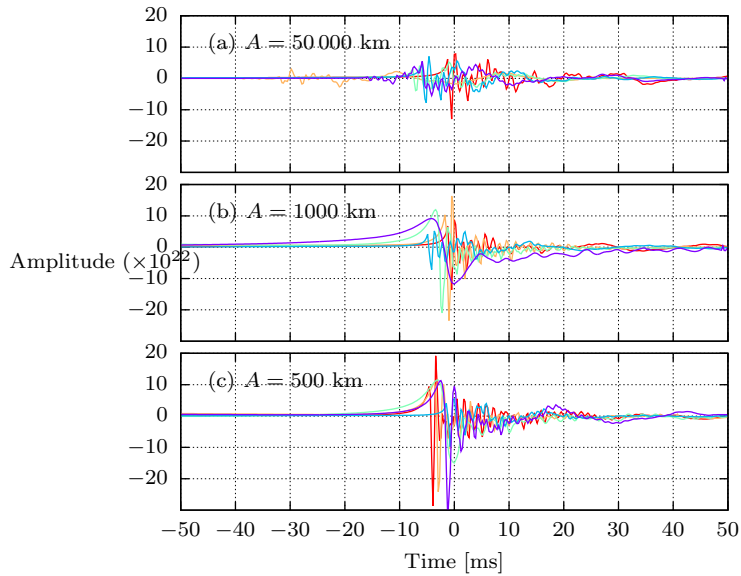}
    \caption{Comparison of waveforms by the values of $A$,
               differentiality of rotation of the initial configuration.
             Five waveforms are chosen at random for each of
               $A = 50\,000$ km (upper panel), $1000$ km (middle panel) and $500$ km (lower panel).}
    \label{fig:SNwaveformsA}
  \end{figure}

  As examples,
    five waveforms for each of $A = 50\,000$, $1000$, and $500$ km are plotted in \fref{fig:SNwaveformsA}.
  It indicates that the amplitude of the gravitational wave from the supernova is high
    if differential rotation of the initial configuration is strong, or $A$ is small.
  It is one of reasons why the efficiency of the EAM largely depends on $A$.
  
\section{Discussion}\label{sec:discussion}
  
\begin{figure}[tbp]
    \centering
    \input{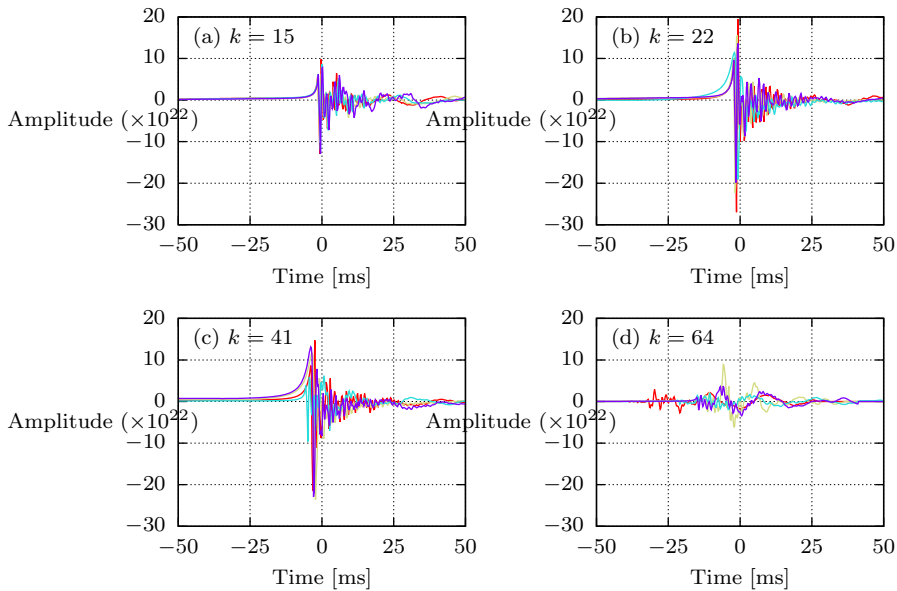}
    \caption{Comparison between waveforms of type k15, k22, k41 and k64.
             Four waveforms are chosen at random for each type.}
    \label{fig:SNwaveforms}
  \end{figure}

\begin{figure}[tbp]
    \centering
    \input{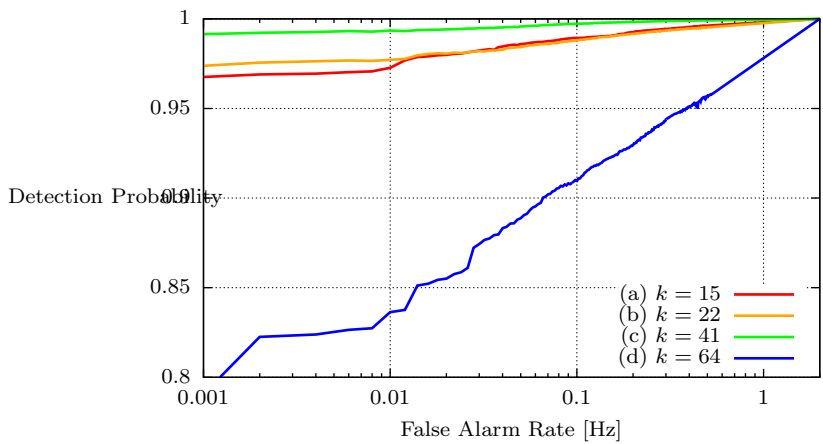}
    \caption{ROC curves for type k15 (red), k22 (orange), k41 (green) and k64 (blue).}
    \label{fig:SNROCk}
  \end{figure}

  The optimal window size depends on the waveform as \fref{fig:SNkoptimal} shows
    and thus the efficiency of the EAM can be improved if the multi-window-size EAM is applied as \fref{fig:SNROCparallel} shows.
  In order to apply the EAM for detecting a specific target,
    it is important to investigate why the optimal window size depends on the waveform.
  In each panel of \fref{fig:SNwaveforms},
    we plotted four of waveforms that have the same value of the optimal window size,
      namely, $k = 15$ (type k15), $k = 22$ (k22), $k = 41$ (k41), and $k = 64$ (k64),
    and the ROC curves of these types are shown in \fref{fig:SNROCk}.
  \Tref{tab:countAk} shows the breakdown by the value of $A$ of each type.
  
  The waves of types k22 and k41 have similar shapes,
    that is, one sharp peak with little transient followed by a short ringdown.
  The optimal window size of k41 is,
    however, larger than that of k22,
      since the peak and the beginning part of the ringdown of k41 are decomposed into the single IMF by the EMD,
      while those of k22 are decomposed into different IMFs.
  The waves of k15 have a short peak but the amplitude is low,
    and hence the optimal window size will be small to compensate for the low amplitude
      with a contribution from the short peak maximally.
  Since the waves of k64 have no significant peak and the amplitude is low,
    the detection probability is much lower than other types as \fref{fig:SNROCk} shows.
  As shown in \tref{tab:countAk},
    the parameter $A$ of all the member of this type except few exceptions is $50\,000$ km,
    that is, they are almost uniformly rotating at the initial time.
  It means that uniform rotation causes the low amplitude
    and thus leads to the difficulty in detecting waves.
  The large window size is required for averaging out the noise to reduce false alerms,
    rather than enhancing the efficiency.
  
  Since only amplitude variations of the IMFs are considered at present,
    it is difficult to detect signals whose amplitude is low throughout even though the duration is long.
  
  \begin{table}[tbp]
    \caption{The number of waveforms that have the optimal window size $k = 15, 21, 41$, and $64$.
             The numbers are listed dividedly according to the value of $A$.}
    \label{tab:countAk}
    \lineup
    \begin{indented}
    \item[]
      \begin{tabular}{cccc}
      \br
        optimal $k$   &  $A = 50\,000$ km  &  $A = 1000$ km  &  $A = 500$ km
      \\
      \mr
                 15  &             {\0}7  &           {\0}5  &         {\0}5
      \\
                 22  &             {\0}4  &              11  &         {\0}4
      \\
                 41  &             {\0}4  &              17  &            38
      \\
                 64  &                20  &           {\0}1  &         {\0}1
      \\
      \br
    \end{tabular}
    \end{indented}
  \end{table}

\section{Summary} \label{sec:summary}
  We validated a possibility of a new amplitude-based approach
    for the detection of gravitational wave bursts
    with the Hilbert-Huang transform.
  Our proposed method, called {\it Excess Amplitude Method},
    will detect an event when the maximum value of all the maximum value
      in each intrinsic mode function of an observed data is greater than a predetermined threshold value.
  Using the simulated time-series noise data and waveforms from rotating core-collapse supernovae at $30$ kpc,
    we performed the simulation to evaluate the performance of our proposed method and found
      that the detection probability reaches 0.94 without false alerms,
      which corresponds to the false alarm rate $< 10^{-3}$ Hz in our simulation.
  
  At the present,
    we do not take variations of the instantaneous frequency of the IMFs into account,
    but it must provide the useful information to distinguish signals from noise.
  We will include the information of the instantaneous frequency
    in the next version of our method.
  
  We used simulated Gaussian noise of Advanced LIGO in this paper,
    but the noise of real laser interferometer detectors show non-Gaussianity and non-stationarity.
  We will apply our method to real laser interferometer data
    and compare the results with other detection methods such as the excess power method soon.
  In addition,
    we are planning to consider the coincidence analysis based on our method.

\ack
  This work was in part supported by MEXT Grant-in-Aid for Scientific Research on Innovative Areas
    ``New Developments in Astrophysics Through Multi-Messenger Observations of Gravitational Wave Sources''
    (Grant Number 24103005).
  This work was also supported in part by JSPS Grant-in-Aid for Scientific Research (C) (Grant Number 15K05071; K.~Oohara),
    by JSPS Grant-in-Aid for Young Scientists (B) (Grant Number 26800129; H.~Takahashi).

\section*{References}


\begin{thebibliography}{10}
  \expandafter\ifx\csname url\endcsname\relax
    \def\url#1{{\tt #1}}\fi
  \expandafter\ifx\csname urlprefix\endcsname\relax\def\urlprefix{URL }\fi
  \providecommand{\eprint}[2][]{\url{#2}}
  
  \bibitem{aLIGO2015}
  Aasi J {\it et~al.\/} (The LIGO Scientific Collaboration) 2015 {\it Class. Quantum Grav.\/} {\bf 32} 074001
  
  \bibitem{AdvVirgo2015}
  Acernese F {\it et~al.\/} 2015 {\it Class. Quantum Grav.\/} {\bf 32} 024001
  
  \bibitem{KAGRA2013}
  Aso Y, Michimura Y, Somiya K, Ando M, Miyakawa O, Sekiguchi T, Tatsumi D and Yamamoto H (The KAGRA Collaboration) 2013 {\it Phys. Rev.\/} D {\bf 88} 043007
  
  \bibitem{ref:cbc}
  Abadie J {\it et~al.\/} 2010 {\it Class. Quantum Grav.\/} {\bf 27} 173001
  
  \bibitem{AndersonPRD2001}
  Anderson W~G, Brady P~R, Creighton J~D~E and Flanagan E~E 2001 {\it Phys. Rev.\/} D {\bf 63} 042003
  
  \bibitem{ref:klimenko2008}
  Klimenko S, Yakushin I, Mercer A, Mitselmakher G 2008 {\it Class. Quantum Grav.} {\bf 25} 114029
  
  \bibitem{ref:sutton2010}
  Sutton P~J {\it et~al.\/} 2010 {\it New Journal of Physics} {\bf 12} 053034
  
  \bibitem{ref:thrane2011}
  Thrane E.{\it et~al.\/} 2011 {\it Phys. Rev.} D {\bf 83} 083004
  
  \bibitem{ref:lynch2015}
  Lynch R, Vitale S, Essick R, Katsavounidis E arXiv:1511.05955
  
  \bibitem{ref:klimenko2015}
  Klimenko S {\it et~al.\/} arXiv:1511.05999
  
  \bibitem{ref:cornish2015}
  Cornish N, Litternberg T, 2015 {\it Class.\ Quantum Grav.\/} {\bf 32} 135012
  
  \bibitem{ref:LIGOPRL2016}
  Abbott B.{\it et~al.\/} (LIGO Scientific Collaboration and Virgo Collaboration) 2016
  {\it Phys.\ Rev.\ Lett.\/}  {\bf 116}, 061102
  
  \bibitem{ref:LIGOburst2016}
  Abbott B.{\it et~al.\/} (LIGO Scientific Collabpration and Virgo Collaboration) 2016
  LIGO P1500229, https://dcc.ligo.org/P1500229/public
  
  \bibitem{HuangPRSA1998}
  Huang N~E, Shen Z, Long S~R, Wu M~C, Shih H~H, Zheng Q, Yen N~C, Tung C~C
    and Liu H~H 1998 {\it Proc. R. Soc.\/} A {\bf 454} 903--995
  
  \bibitem{Jordan2007}
  Camp J~B, Cannizzo J~K and Numata K 2007 {\it Phys. Rev.\/} D {\bf 75} 061101
  
  \bibitem{StroeerPRD2009}
  Stroeer A, Cannizzo J~K and Camp J~B 2009 {\it Phys. Rev.\/} D {\bf 79} 124022
  
  \bibitem{Takahashi2013}
  Takahashi H, Oohara K, Kaneyama M, Hiranuma Y and Camp J~B 2013 {\it
    Advances in Adaptive Data Analysis\/} {\bf 5} 1350010
  
  \bibitem{ref:kaneyama_sub}
  Kaneyama M, Oohara K, Takahashi H, Sekiguchi Y, Tagoshi H, Shibata M, submitted to {\it Phys. Rev.\/} D
  
  \bibitem{FlandrinIEEE2004}
  Flandrin P, Rilling G and Gon\c{c}alv\'{e}s P 2004 {\it IEEE Signal Process. Lett.\/} {\bf 11} 112--114
  
  \bibitem{Dimmelmeier2008}
  Dimmelmeier H, Ott C~D, Marek A and Janka H~T 2008 {\it Phys. Rev.\/} D {\bf 78} 064056
    
  \bibitem{ref:aligonoisecurve}
  https://dcc.ligo.org/cgi-bin/DocDB/ShowDocument?docid=2974.
  
  \bibitem{ShenPTP1998}
  Shen H, Toki H, Oyamatsu K and Sumiyoshi K 1998 {\it Progr. Theoret. Phys.\/} {\bf 100} 1013
  
  \bibitem{LattimerNPA1991}
  Lattimer J~M and Swesty F~D 1991 {\it Nucl. Phys.\/} A {\bf 535} 331--376
  
  \bibitem{TheoryOfLinearPrediction}
  Vaidyanathan P~P 2008 {\it The Theory of Linear Prediction\/} (Morgan \& Claypool)
  
  \bibitem{CuocoCQG2001}
  Cuoco E, Calamai G, Fabbroni L, Losurdo G, Mazzoni M, Stanga R and Vetrano F 2001 {\it Class. Quantum Grav.\/} {\bf 18} 1727
  
  \bibitem{McNabbCQG2004}
  McNabb J~W~C, Ashley M, Finn L~S, Rotthoff E, Stuver A, Summerscales T, Sutton P, Tibbits M, Thorne K and Zaleski K 2004 {\it Class. Quantum Grav.\/} {\bf 21} S1705--S1710
  
  \bibitem{ChatterjiCQG2004}
  Chatterji S, Blackburn L, Martin G and Katsavounidis E 2004 {\it Class. Quantum Grav.\/} {\bf 21} S1809--S1818
  
  \bibitem{Cohen1995}
  Cohen L 1995 {\it Time-Frequency Analysis\/} (Prentice Hall)

  \end{thebibliography}
\end{document}